\documentclass{PoS}

\usepackage{cite}

\usepackage{amssymb}

\title{$p_T$-fluctuations and multiparticle clusters in heavy-ion 
collisions\thanks{Supported in part by the Polish Ministry of 
Education and Science grants 2P03B~02828, Funda\c{c}\~{a}o para
a Ci\^{e}nca e a Tecnologia, POCI/FP/63930/2005, POCI/FP/63412/2005, PRAXIS
XXI/BCC/429/94, and GRICES.} }

\ShortTitle{$p_T$-fluctuations and multiparticle clusters in heavy-ion 
collisions }

\author{\speaker{Wojciech Broniowski}$^{ab}$, Piotr Bo\.zek$^{bc}$, Wojciech Florkowski$^{ab}$, 
Brigitte Hiller$^d$ \\
        \llap{$^a$}{Institute of Physics, \'Swi\c{e}tokrzyska Academy,
ul.~\'Swi\c{e}tokrzyska 15, PL-25406~Kielce, Poland} \\
        \llap{$^b$}{The H. Niewodnicza\'nski Institute of Nuclear Physics, 
Polish Academy of Sciences, \\ 
PL-31342 Krak\'ow, Poland}\\
        \llap{$^c$}{Institute of Physics, Rzesz\`ow University, PL-35959 Rzeszów, Poland}\\
        \llap{$^d$}{Centro de F{\'{i}}sica Te\'orica, Departamento de F{\'{i}}sica, 
University of Coimbra, P-3004-516 Portugal}\\
    E-mail: \email{Wojciech.Broniowski@ifj.edu.pl}, 
            \email{Piotr.Bozek@ifj.edu.pl},
            \email{Wojciech.Florkowski@ifj.edu.pl},
            \email{brigitte@teor.fis.uc.pt} }

\abstract{We investigate the centrality dependence of the $p_T$-correlations 
in the event-by-event analysis of 
relativistic heavy-ion collisions at RHIC made recently by the PHENIX and STAR 
Collaborations. We notice that $\sigma^2_{\rm dyn}$ scales 
to a very good accuracy with the inverse number of the produced particles, $n$.
This scaling can be naturally explained by formation of clusters. 
We discuss the nature of clusters and provide numerical estimates 
of correlations coming from resonance decays, which are tiny, and a model 
where particle are emitted from local thermalized sources moving at the same collective
velocity. This "lumped cluster'' model can explain the data when on the average 6-15 particles are 
contained in a cluster. We also point out simple relations of the popular correlation 
measures to the covariance, which arise under the assumptions (well holding at RHIC) that the  
distributions are sharply peaked in $n$ and that the dynamical compared to statistical 
fluctuations are small.}

\FullConference{Correlations and Fluctuations in Relativistic Nuclear Collisions\\
                 July 7-9 2006\\
                 Florence, Italy}

\PACS{25.75.-q, 25.75.Gz, 24.60.-k}

\begin{document}

\section{Introduction}

We have had very interesting presentations on event-by-event fluctuations
during this workshop (see other contributions to these proceedings). 
This work is based on our recent paper \cite{Broniowski:2005ae}
and explores from a theoretical viewpoint the 
correlation data published 
by the PHENIX \cite{Adcox:2002pa,Adler:2003xq}, STAR \cite{Adams:2003uw,Adams:2005ka}, and CERES
\cite{ceres} 
Collaborations. These measurements contribute to the previously accumulated 
vast knowledge on 
event-by-event fluctuations in ultra-relativistic heavy-ion collisions 
\cite{these,MIT,Gazdzicki:1992ri,Stodolsky:1995ds,Shuryak:1997yj,%
Mrowczynski:1997kz,Stephanov:1999zu,Voloshin:1999yf,%
Appelshauser:1999ft,ceres,Utyuzh:1999iu,Korus:2001au,%
Baym:1999up,Bialas:1999tv,portugal,port2,port3,Asakawa:2000wh,Heiselberg:2000fk,Utyuzh:2001up,%
Pruneau:2002yf,Jeon:2003gk,Gavin:2003cb,Liu:2003jf,%
Abdel-Aziz:2005wc,aziz2}. 

We bring up some basic and rather striking
facts manifest in the data of Refs.~\cite{Adcox:2002pa,Adams:2005ka} and 
discuss their relevance to the {\em cluster picture} of correlations. 
Let us begin with the PHENIX measurement \cite{Adcox:2002pa} of the event-by-event 
fluctuations of the transverse momentum at $\sqrt{s_{NN}}=130$~GeV. 
In order to simplify the notation, $p$ is used generically 
to denote the transverse momentum, $|\vec{p}_T|$ and $p_i$ is the value of 
$p$ for the $i$th particle. Finally, 
$M = \sum_{i=1}^n p_i/n $ is 
the average transverse momentum in an event of multiplicity $n$. 
The statistical treatment of the $M$ variable has been presented in detail in 
the contribution of the speaker to the discussion session (see {\em ``Round Table Discussion:
Correlations and Fluctuations in Nuclear Collisions''} in these proceedings \cite{disc}).
For the reader's convenience this material is inluded here in Appendix~A.

\section{A look at the data \label{sec:look}}

The PHENIX results of Ref.~\cite{Adcox:2002pa} are reminded in Table~\ref{tab:data}.
\begin{table}[b]
\caption{\label{tab:data} 
Data for the event-by-event fluctuations in the transverse momentum. Upper rows: 
the PHENIX measurement at $\sqrt{s_{NN}} =130$~GeV  \cite{Adcox:2002pa}. 
Bottom rows: the mixed-event results. }
\medskip
\begin{center}
\begin{tabular}{|l|c|c|c|c|}
\hline
centrality & 0-5\% & 0-10\% & 10-20\% & 20-30\% \\
\hline
$\bar n$ & 59.6 & 53.9 & 36.6 & 25.0 \\
$\sigma_n$ & 10.8 & 12.2 & 10.2 & 7.8 \\
$\bar M$ & 523 & 523 & 523 & 520 \\
$\sigma_p$ & 290 & 290 & 290 & 289 \\
$\sigma_M$ & 38.6 & 41.1 & 49.8 & 61.1 \\
\hline
$\bar M^{\rm mix}$ & 523 & 523 & 523 & 520 \\
$\sigma_M^{\rm mix}$ & 37.8 & 40.3 & 48.8 & 60.0 \\
\hline
\end{tabular}
\end{center}
\label{tab:phen}
\end{table}
Several important features of the data are immediately seen: firstly, the 
quantities $\bar M$, the average $p$ in a given centrality class, and $\sigma_p$, the inclusive
standard deviation of $p$, are practically constant in the 
reported ``fiducial centrality range'' $c=0-30$\%.
\begin{eqnarray}
\bar M  = \hbox{const.}\,, \;\;\;\;\;  \sigma_p = \hbox{const.}
~~(\rm{at~low}~c).  \label{const}
\end{eqnarray}
This allows us to replace the average momentum at each $n$ by $\bar M$ and the variance at each $n$ by $\sigma_p$ 
\cite{Broniowski:2005ae}. For distributions sharply peaked in the $n$ variable we derive in \cite{disc} the
following formula:
\begin{equation}
\sigma_M^{\rm mix} \simeq \sigma_p/\sqrt{\bar n}. \label{eq:varmix}
\end{equation}
The derivation involves only the statistics and the assumption of sharp distributions, which is
sufficiently well satisfied at RHIC. Corrections to Eq.~(\ref{eq:varmix}) appearing for 
broader distributions can be obtained. They involve higher moments of $n$, for instance 
$\sigma^2(n)/{\bar n}^2$. 
Table~\ref{tab1} shows how well (\ref{eq:varmix}) works -- at the level of 1-2\%, which is no
wonder, just the statistical fact. 

\begin{table}[tb]
\caption{ \label{tab1} Verification of the statistical result (2.2) 
holding for sharp distributions. } 
\begin{center}
\begin{tabular}{|l|c|c|c|c|}
\hline
centrality & 0-5\% & 0-10\% & 10-20\% & 20-30\% \\
\hline
$\sigma_M^{\rm mix}$ & { 37.8} & { 40.3} & { 48.8} & { 60.0} \\
$\sigma_p /\sqrt{\bar n}$ 
& { 37.6} & { 39.5} & { 47.9} & { 59.0} \\
\hline
\end{tabular}
\end{center}
\end{table}

\section{The cluster scaling \label{sec:cluster}} 

Next, we look at the difference of the measured and mixed-event variances of $M$, which
is a measure of dynamical fluctuations, 
\begin{eqnarray}
\sigma_{\rm dyn}^2 \equiv \sigma_M^2 - \sigma_M^{2, {\rm mix}}.
\label{dyn} 
\end{eqnarray}
Table \ref{tab2} shows our main observation, namely, 
that the following scaling holds:%
\begin{eqnarray}
\sigma_{\rm dyn} \sim \frac{1}{\sqrt{\bar n}}. 
\label{scaledyn} 
\end{eqnarray}
It works at the accuracy level of 2\% in the available centrality range, as can be inferred 
from the approximate constancy of numbers displayed in Table~\ref{tab2}. 
\begin{table}[tb]
\caption{\label{tab2} Verification of the scaling result (3.2), seen in the 
approximate constancy of $\sigma_{\rm dyn} \sqrt{\bar n}$.}
\begin{center}
\begin{tabular}{|l|c|c|c|c|}
\hline
centrality & 0-5\% & 0-10\% & 10-20\% & 20-30\% \\
\hline
$\sigma_{\rm dyn} \sqrt{ \bar n}$ & {$60.3 \pm 1.6$} & {$59.2 \pm 1.5$} & 
{$59.8 \pm 1.2$} & {$57.7 \pm 1.1$} \\
\hline
\end{tabular}
\end{center}
\end{table}
Elementary statistical considerations shown in \cite{disc} lead to the result
\begin{eqnarray}
\sigma_{\rm dyn}^2 =
\sum_n \frac{P(n)}{n^2}  \sum_{i,j=1, j \neq i}^n  {\rm cov}_{n}(p_i,p_j) 
\simeq \frac{1}{\bar n^2} \sum_{i,j=1, j \neq i}^{\bar n} 
 {\rm cov}_{\bar n}(p_i,p_j), 
\label{dyn2}
\end{eqnarray}
where in the second equality we have used the feature of sharp distributions in $n$.
The quantity ${\rm cov}_{n}(p_i,p_j)$ is the covariance of momenta $p_i$ and $p_j$ 
in events of multiplicity $n$.  

\begin{figure}[tb]
\begin{center}
\includegraphics[angle=0,width=8.5cm]{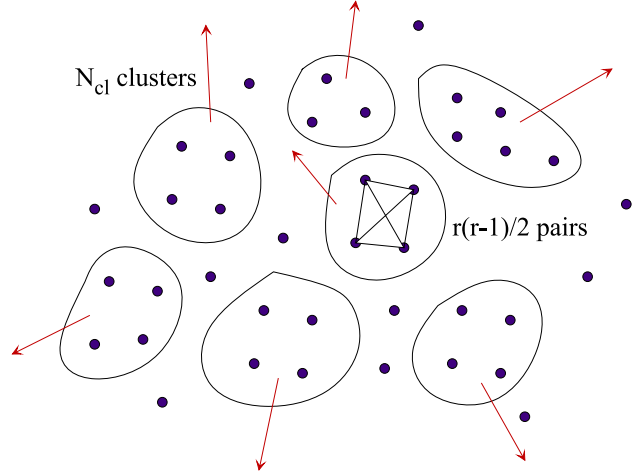}
\end{center}
\caption{\label{fig:clust} The cluster picture of correlations 
consistent with scaling (1.4).}
\end{figure}

Now we come to the physics discussion.  The scaling (\ref{scaledyn}) imposes severe 
constraints on the physics nature of the covariance term. For instance, if all 
particles were correlated to each other, then
$\sum_{i,j=1, j \neq i}^n {\rm cov}_{\bar n}(p_i,p_j)$ would be proportional 
to the number of all pairs, and $\sigma_{\rm dyn}$ would be independent of 
$\bar n$. Thus we see that the combinatorics is truncated -- clearly not all pairs
are correlated, or a finite correlation length develops.  
A natural explanation of the scaling (\ref{scaledyn}) comes from the 
cluster model, depicted in Fig.~\ref{fig:clust}. The system is assumed to have $N_{\rm cl}$
clusters, each containing (on the average) $\langle r \rangle$ particles. 
The particles are correlated if and only if they belong to the 
same cluster, where the average covariance per pair is $2\,{\rm cov}^\ast$. The number of 
correlated pairs within a cluster is $r(r-1)/2$. Some particles may be unclustered,
hence the ratio of clustered to all particles is 
$\langle N_{\rm cl}\rangle \langle  r  \rangle /{\bar n} = \alpha$. If all particles are clustered
then $\alpha=1$. With these assumptions Eq.~(\ref{dyn}) becomes
\begin{eqnarray}
\sigma_{\rm dyn}^2 = \frac{\alpha \langle r(r-1) \rangle }{\langle r \rangle \bar n} {\rm cov}^\ast
= \frac{\alpha r^\ast}{\bar n} {\rm cov}^\ast, 
\label{cluster}
\end{eqnarray}
where we have introduced $r^\ast=\langle r(r-1) \rangle / \langle r \rangle$, the ratio of the average number of 
pairs in the cluster to the average multiplicity of the cluster. For a fixed number of particles in each cluster
we have $r^\ast=\langle r \rangle-1$, for
the Poisson distribution $r^\ast=\langle r \rangle$, while for wider distributions 
$r^\ast > \langle r \rangle$. 
Equation (\ref{cluster}) complies to the scaling (\ref{scaledyn}) as long as the product
$\alpha r^\ast {\rm cov^\ast}$ does not depend on $\bar n$ 
(in the fiducial centrality range).
This is the basic physics constraint that follows from the data. 
For the reasons discussed above it is appropriate to 
term the scaling (\ref{scaledyn}) the {\em cluster scaling}.

\section{Cluster scaling with other correlation measures}

There are numerous quantities used as measure of the event-by-event 
fluctuations. Fortunately, as shown in \cite{disc}, for sharp distributions and 
for the case where the dynamical correlations are small compared to the 
statistical correlations, $\sigma_{\rm dyn} \ll \sigma_M$, {\em all the popular 
measures are simply related to each other}, or to the {\em covariance}, which is the basic 
quantity designed to describe correlations. We have
\begin{eqnarray}
\Sigma^2_{p_T} &\equiv& {\sigma^2_{\rm dyn}}/{\overline{p}^2} 
\simeq \frac{\sum_{i \neq j} {\rm cov}_{\overline{n}}(p_i,p_j)}{{\bar p}^2 {\bar n}^2} ,\nonumber \\
F_{p_T} &\equiv& \frac{\sigma_M}{\sigma^{{\rm mix}}_M } -1=\sqrt{1+\frac{\sigma^2_{\rm dyn}}
{\sigma^{2,{\rm mix}}_M }}-1 \simeq 
\frac{1}{2} \frac{\sigma^2_{\rm dyn} }{\sigma^{2,{\rm mix}}_M} \simeq 
\frac{\sum_{i \neq j} {\rm cov}_{\overline{n}}(p_i,p_j)}{2 {\bar n} \sigma_p^2}, \nonumber \\ 
\Phi_{p_T} &\equiv& \sqrt{\frac{\sigma_S^2}{\overline{n}}}-
\sigma_p \simeq \frac{\sum_{i \neq j} {\rm cov}_{\overline{n}}(p_i, p_j)} 
{2 \overline{n}\sigma_p}, \nonumber \\
\langle \Delta p_i \Delta p_j \rangle &\equiv& \sum_{i \neq j} {\rm cov}_{\overline{n}}(p_i,p_j),
\label{eq:relation}
\end{eqnarray}
where in the definition of $\Phi_{p_T}$ one uses $S_n= p_1+\dots+p_n$. 
The last equality states the fact that the STAR measure $\Delta p_i \Delta p_j$ is just the statistical
estimator of the covariance, see \cite{Broniowski:2005ae}. 
The cluster scaling thus manifests itself for the various measures in the following manner:
\begin{eqnarray}
\Sigma_{p_T} &\sim& \frac{1}{\sqrt{\bar n}} ,\nonumber \\
F_{p_T} &\sim& 1, \nonumber \\ 
\Phi_{p_T} &\sim& 1, \nonumber \\
\bar n \langle \Delta p_i \Delta p_j \rangle &\sim& 1. \label{eq:relscale}
\end{eqnarray}
Since this is an important point, we stress again: 
since at RHIC the distributions are 1)~sufficiently sharply peaked in $n$ and 2)~the dynamical 
fluctuations are small compared to the statistical fluctuations, 
all popular measures of event-by-event 
fluctuations are all proportional to each other and to the covariance. One may pass from one measure
to another without difficulty. If conditions 1) and 2) are not satisfied, as may be the case at 
lower energies, periferal collisions, {\it etc.}, then of course the measures remain 
different and some may be better tuned for certain analyses.
Full information on correlations could be acquired 
by simply evaluating the covariance $\sum_{i \neq j} {\rm cov}_n$ separately for each $n$. 
Even if 1) or 2) are relaxed, 
the correlation measures are still related to the 
sum of the weighted covariances at various $n$, with the weights dependent on the particular measure. 
The necessary formulas can be very straightforwardly derived along the lines of \cite{disc}.
We urge that such a study of the dependence of covariance on 
$n$ be made on data with sufficiently large samples.  

\section{Cluster scaling in various experiments}

Now let us have a look at various experiments. The PHENIX results at $\sqrt{s_{NN}}=130$~GeV 
\cite{Adcox:2002pa} discussed in Sect.~\ref{sec:cluster} comply 
nicely to the cluster scaling. The same is true of 
the STAR data (see Fig.~3 of \cite{Adams:2005ka}) at $\sqrt{s_{NN}}=200$~GeV and 62~GeV, where 
the quantity $d n/d\eta \langle \Delta p_i \Delta p_j \rangle$ flattens out for large centralities. 
For $\sqrt{s_{NN}}=130$~GeV and 20~GeV the STAR data seems to somewhat depart from the scaling, with 
sizeable error bars for 130~GeV. The PHENIX data at $\sqrt{s_{NN}}=200$~GeV shows non-monotonic
behavior \cite{Adler:2003xq}. The STAR data at $\sqrt{s_{NN}}=200$~GeV also shows non-monotonic
behavior at large centralities, but only for the analysis where all produced 
particles are taken into account. When the analysis is constrained to the in-plane or out-of-plane
particles, then the cluster scaling is satisfied remarkably well (see 
Paul Sorensen's talk showing the STAR preliminary data). This is a very intriguing phenomenon that needs to be understood.
The CERES data \cite{Appelshauser:1999ft}, using the $\Sigma_{p_T}$ variable, 
also complies to the cluster scaling within the error bars. Recall that according to (\ref{eq:relscale}) we look for
$\Sigma_{p_T} \sim {1}/\sqrt{\bar n}$. In conclusion, the cluster scaling is seen in some measurements
to a remarkable accuracy, while is other non-monotonous behaviour is apparent. The situation requires 
careful clarification. 

\section{Strength of correlations}

Before performing the analysis of clusters in a more quantitative manner 
we need to consider the effects of acceptance and the detector efficiency. This is particularly
important in the event-by-event analysis, since the experiments select
particles { with very clearly identified tracks}, hence the detector efficiency $a$ is reduced. 
The number of observed particles is proportional to $a$, 
and the number of pairs contributing to the covariance is proportional to $a^2$. 
Thus Eq.~(\ref{cluster}) may be rewritten as 
\begin{equation}
\sigma_{\rm dyn}^2 = \frac{r^\ast}{\bar n}_{\rm full}
{\rm cov}^\ast = a \frac{r^\ast}{\bar n}_{\! \rm obs} {\rm cov}^\ast,
\end{equation}
where ``full'' denotes all particles (that would be observed with 100\% efficiency), while ``obs''
stands for the actually observed multiplicity of particles. Thus 
\begin{eqnarray}
{\rm cov}^\ast = \sigma_{\rm dyn}^2 \frac{\bar n_{\! \rm obs}}{a r^\ast}. 
\label{clust2}
\end{eqnarray}
Our estimate for $a$ in the PHENIX experiment is of the order of
10\%, which together with the numbers of Table \ref{tab:data} gives  
\begin{eqnarray}
{\rm cov}^\ast \simeq \frac{0.035~{\rm GeV}^2}{r^\ast}. 
\label{clust3}
\end{eqnarray}
In the considered problem the coefficient $0.035~{\rm GeV}^2$ is not a small
number when compared to  the natural
scale set by the variance $\sigma_p^2 \simeq 0.08~{\rm GeV}^2$ (we recall that 
$\mid {\rm cov}^\ast \mid \le \sigma_p^2$). 

Very similar quantitative conclusions are reached with the STAR data. 
Taking the values from Table~I of Ref.~\cite{Adams:2005ka} and guessing \mbox{$a=0.75$} we find 
$ {\rm cov}^\ast r^\ast=0.058, 0.043, 0.035, 0.014~{\rm GeV}^2$ for 
$\sqrt{s_{NN}}=200, 130, 62$ and 20 GeV,
respectively. The value at 130~GeV is close to the PHENIX value (\ref{clust3}). 
Interestingly, we note a 
significant beam-energy dependence, with ${\rm cov}^\ast r^\ast$ increasing with $\sqrt{s_{NN}}$. 
This may be due to the increase of the covariance per correlated pair with the 
increasing energy, and/or an increase of the number of particles within a cluster. 

Using the above numbers we find that for 
$r^\ast=1$ (for instance the case where all clusters have just two particles) 
the value of ${\rm cov}^\ast$ assumes almost a half of the maximum possible value, 
$\sigma_p^2$. 
This is very unlikely, as  dynamical estimates presented below give ${\rm cov}^\ast$ of the order 
at most $0.01~{\rm GeV}^2$. Thus a natural explanation of the values in (\ref{clust3}) is to take 
a significantly larger value of $r^\ast$ -- just put more particles inside the cluster. 
Of course, the higher value, the easier it is to satisfy 
(\ref{clust3}) even with small values of ${\rm cov}^\ast$. 
This scenario will be elaborated in the next section.

\section{The nature of clusters}

Several mechanisms can be brought up to describe the formation of clusters
in the momentum space. The basic physics question concerns {\em the nature of clusters}. 
Here we discuss a few popular scenarios. 

\subsection{Jets}

The most prominently explored mechanism is the formation of
(mini)jets. In Ref.~\cite{Broniowski:2005ae} we have discussed this issue, with the conclusion 
that the explanation of the centrality dependence of the 
$p_T$ fluctuations in terms of jets based solely on scaling arguments is not
conclusive. Any mechanism leading to clusters in $p_T$ would do.
Microscopic realistic estimates of the average number of jets and 
the magnitude of the covariance of particles 
originating from a jet
are necessary here, including the interplay of jets and the medium. 
For the current status of this program the user
is referred to \cite{Liu:2003jf,PHENIX:mitchell}.

\begin{figure}[tb]
\begin{center}
\includegraphics[width=7.5cm,angle=0]{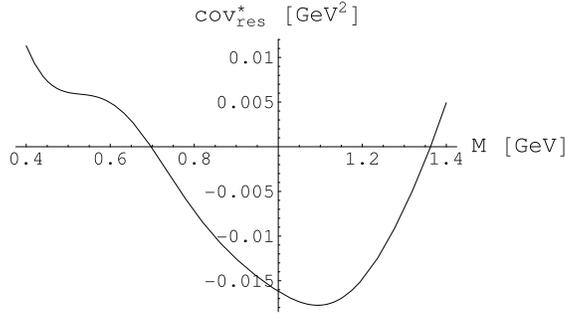}
\end{center}
\caption{The covariance ${\rm cov}^\ast_{\rm res}$ of the 
two pions produced in the decay of the resonance om mass $M$.}
\label{fig:rhoc}
\end{figure}

\subsection{Resonance decays}

Another natural mechanism for momentum correlations is provided by resonance
decays. Recall that resonances are crucial behind the success of the statistical 
approach to particle production in relativistic heavy-ion collisions.
We have made an estimate of this effect in the thermal model of 
Ref.~\cite{Broniowski:2001we,Broniowski:2002nf}. The covariance ${\rm cov}^\ast_{\rm res}$ of the 
two pions produced in a resonance decay is given by the
formula
\begin{eqnarray}
\hspace{-1cm} && {\rm cov}^\ast_{\rm res} 
= {\!\int\! d^3p \!\int {d^3p_1 \over E_{p_1}} \!\int {d^3p_2 \over E_{p_2}} 
\,\delta^{(4)}(p-p_1-p_2)\; C \; {dN_R \over d^3p} 
\left(p_1^T\! -\! \langle p^T \rangle \right) 
\left(p_2^T\! - \!\langle p^T \rangle \right) 
\over
\!\int\! d^3p \int {d^3p_1 \over E_{p_1}} \int {d^3p_2 \over E_{p_2}} 
\, \delta^{(4)}(p-p_1-p_2) \; C \; {dN_R \over d^3p} 
}, 
\label{thermalcov}
\end{eqnarray}
where $dN_R / d^3p$ is the resonance distribution in the momentum space (obtained
from the Cooper-Frye formula as described in Ref. \cite{Bozek:2003qi}), $M$ is the mass of the
resonance,  $p_1$ and $p_2$ are 
the momenta of the emitted particles, $E_p$ is the energy of a particle with momentum $p$,
and the function $C$ represents the appropriate experimental cuts. Here we take $C$ as for the 
PHENIX experiment, with 0.2~GeV~$< p_{1,2}^T <1.5$~GeV, and $|y| \le 0.35$. The average 
momentum of the pion is $\langle p^T \rangle=523$~MeV. The parameters of the model are 
$T=165$~MeV (the universal freeze-out temperature), $\rho_{\rm max}=7.15$~fm (the transverse size 
of the system), and $\tau=7.86$~fm (the proper time at freeze-out). 
The numerical results presented in Fig.~\ref{fig:rhoc} show that for the resonance mass between 
500~MeV and 1.2~GeV the covariance ${\rm cov}^\ast_{\rm res} $ varies between 
0.01~GeV$^2$ at low masses to $-0.018$~GeV$^2$ at $M \sim 1.1$~GeV. These numbers are 
small compared to the scale $\sigma_p^2 \simeq 0.08$~GeV$^2$. Moreover, there
is a  change of sign around $M=700$~MeV and cancellations between contributions of 
various resonances are possible. In fact, a full-fledged simulation with {\tt Therminator} 
\cite{Kisiel:2005hn} revealed a completely 
negligible contribution of resonances to the $p_T$ correlations with model parameters 
appropriate for the STAR and PHENIX experiments. 

\subsection{Lumps of matter}
   
Finally, let us consider a model of momentum correlations which assumes that the particle emission
at the lowest scales occurs from local thermalized sources. We call this picture the 
``lumped clusters'': 
lumps of matter move at some collective velocities, correlating the momenta of particles belonging 
to the same cluster, see Fig.~\ref{fig:clust}. 
Each element of the fluid moves with its collective velocity and emits
particles  with locally thermalized spectra. This picture was put forward as a
mechanism creating correlations in the  charge balance function 
\cite{Bozek:2003qi,Cheng:2004zy}
resulting from   charge conservation within a local source. 
The covariance between particles $i$ and $j$ emitted from a cluster moving with a velocity $u$ is
given by the equation
\begin{equation}
\hspace{-0.75cm} {\rm cov}^\ast(i,j)=\frac{ \int d\Sigma_\mu u^\mu 
\int d^3p_1  (p^T_1-\langle p^T \rangle )
f_i^u(p_1) \int d^3p_2 (p^T_2- \langle p^T \rangle ) f_j^u(p_2)}
 { \int d\Sigma_\mu u^\mu 
\int d^3p_1 f_i^u(p_1) \int d^3p_2
f_j^u(p_2)}, 
\end{equation}
where $f_i^u(p)=(\exp(p \cdot u /T)\pm 1)^{-1}$ is the thermal
distribution in the local reference frame and $d\Sigma_\mu $ denotes integration over the freeze-out hypersurface. 
In this calculation we adjust the average transverse flow velocity $\langle \beta_T \rangle$ at each $T$ such that the 
slope of the pion spectra agrees with the data. Of course, lower $t$ requires higher $\langle \beta_T \rangle$.            
The result turns out to depend strongly on the temperature. 
For the emission of correlated pion pairs one gets 
the results shown in Table~\ref{tab:therm}.
\begin{table}[tb]
\caption{\label{tab:therm} Predictions of the ``lumped cluster'' model. The quantity $T$ is the universal 
freeze-out temperature, $\langle \beta_T \rangle$ is the average transverse flow velocity, ${\rm cov}^\ast_{\pi\pi}$ 
is the average covariance of the pion pair, and $r^\ast$ is an estimate for the average number of charged particles in the 
cluster based on the PHENIX results for $\sqrt{s_{NN}}=130$~GeV.}
\begin{center}
\begin{tabular}{|l||c|c|c|c|c|c|}
\hline
 $T$ [MeV]                           & 10    & 100  & 120  & 140  & 165  & 200  \\
 $\langle \beta_T \rangle$             & 0.94  & 0.72 & 0.69 & 0.58 & 0.49 & 0.31 \\
 $ \sigma_p^2$ [GeV$^2$]             & 0.056 & 0.19 & 0.15 & 0.15 & 0.14 & 0.12 \\
 ${\rm cov}^\ast_{\pi\pi}$ [GeV$^2$] & 0.027 & 0.011& 0.0088& 0.0063& 0.0034& 0.0006 \\
 $r^\ast=\frac{0.035~{\rm GeV}^2}{{\rm cov}^\ast_{\pi\pi}}$ & 1.3 & 3.2 & 4.0 & 5.6 & 10.3 & 58.3 \\
\hline
\end{tabular}
\end{center}
\end{table}
The last row of the table gives the number of particles based on 
formula (\ref{clust3}), obtained for PHENIX at $\sqrt{s_{NN}}=130$~GeV.
We note that for realistic values of  thermal freeze-out
parameters the experimentally estimated value of the covariance cannot be
accounted for, unless the number of charged particles belonging to the same cluster
is at least of the order $4-10$ (assuming the Poisson distribution). For wider distributions
in the variable $r$ a lower number is requested. Thus the number of all particles 
(charged and neutral) belonging to a cluster is estimated as
\begin{equation}
r^\ast_{\rm all} \sim 6-15, \label{qual}
\end{equation}
which is one of our main results.

\section{Dependence on maximum transverse momentum}

\begin{figure}[tb]
\begin{center}
\includegraphics[width=8.5cm]{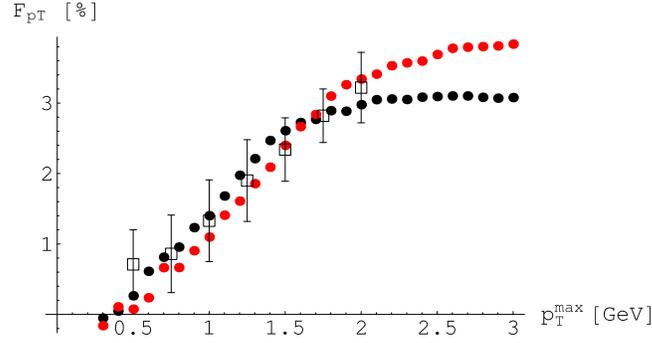}
\end{center}
\caption{Dependence of $F_{p_T}$ on the cut in the transverse momentum. The data points come
from Ref.~\cite{Adler:2003xq}. Red points show the model calculation with $T=165$~MeV, while 
the black points are for $T=130$~MeV.}
\label{fig:fpt}
\end{figure}

An interesting result is obtained when the upper limit of integration 
in the transverse momentum, $p_T^{\rm max}$, is imposed. Figure~\ref{fig:fpt} 
shows the dependence of $F_{p_T}$ on $p_T^{\rm max}$ and compares the result to 
the PHENIX data. The red points show the model calculation with $T=165$~MeV and $a r^\ast=2.1$, while 
the black points are for $T=130$~MeV and $a r^\ast=2.9$. The value of  $a r^\ast$ is tuned such that 
the model describes the data well. For realistic values of the acceptance $a$ we get values of $r^\ast$
of the same order as in estimate (\ref{qual}). 

\begin{figure}[tb]
\begin{center}
\includegraphics[width=14.5cm,angle=0]{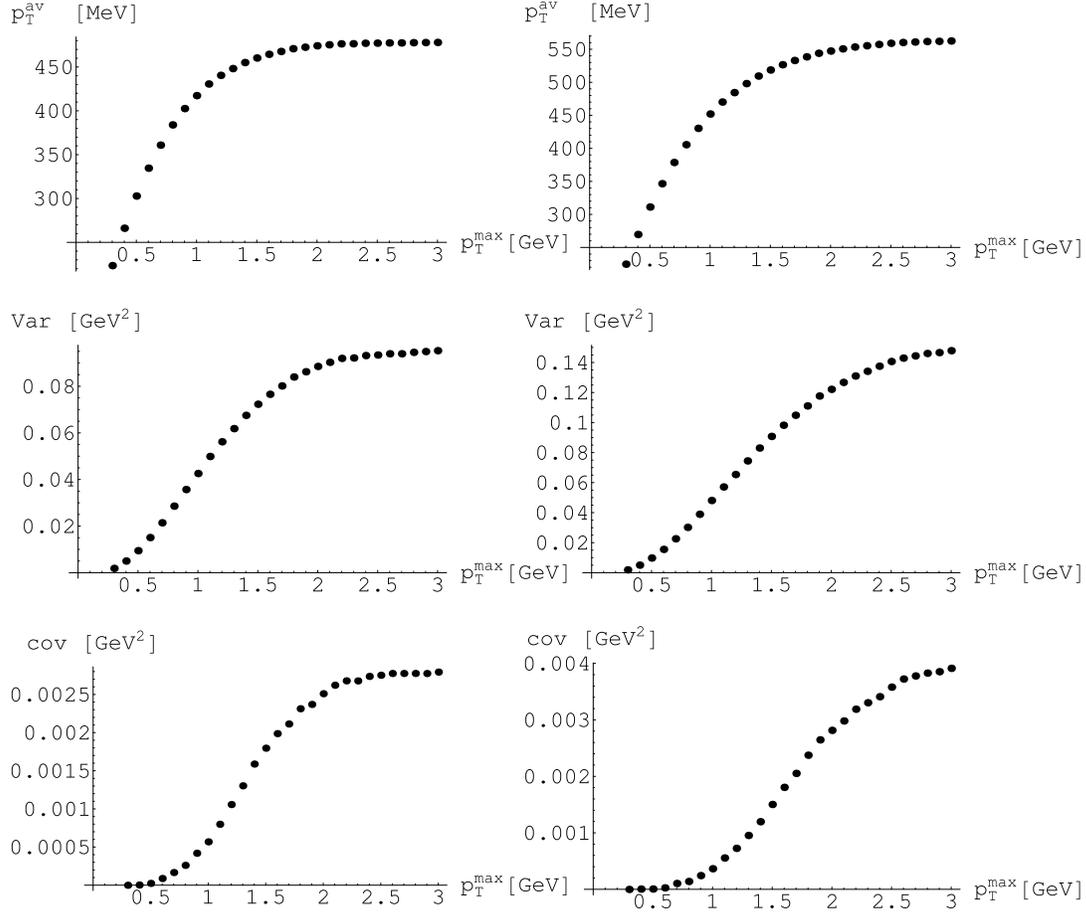}
\end{center}
\caption{The dependence of the average momentum (top), its variance (middle),
and the covariance (bottom) resulting from the lumped cluster model for $T=130$~MeV (left) and
$T=165$~MeV (right) on the cut-off in the transverse momentum, $p_T^{\rm max}$. }
\label{fig:tablo}
\end{figure}

A feature of our model is a saturation of $F_{p_T}$ at 
large values of $p_T^{\rm max}$. This saturation is a consequence of the thermal model. 
As a matter of fact,   
all relevant quantities saturate when integrated over $p_T$. 
Unfortunately, the experimental data do not extend to the saturation region.
What is somewhat surprising at the first
sight is the rather large value of $p_T^{\rm max}$ needed for saturation: 
about 2~GeV for $T=130$~MeV and  
about 3~GeV for $T=165$~MeV. Figure \ref{fig:tablo} shows the results for the average momentum, 
its variance,
and the covariance resulting from the lumped cluster model. We note that all these quantities
saturate at rather large values of $p_T^{\rm max}$. In that sense in the present context the ``soft'' 
thermal physics reaches transverse momenta up to 2-3~GeV! 

\section{Conclusion}

We list our main results:

\begin{enumerate}
\item The {\em cluster scaling} of $\sigma^2_{\rm dyn}$ with $1/{\bar n}$
in the fiducial centrality range 0-30\% at PHENIX
points at the {cluster picture} of the fireball. 
\item Similar scaling can be also seen at STAR \cite{Adams:2003uw,Adams:2005ka} (see also Paul Sorensen's talk)
and at CERES \cite{ceres}.
\item Scaling can be found with various correlation measures, 
which in the limit of sharp distributions and small dynamical compared 
to statistical fluctuations become equivalent and related in simple ways 
to the covariance, as discussed in \cite{disc}.
\item The clusters may a priori originate from
very different physics: (mini)jets, droplets of fluid formed in the explosive scenario of the collision, 
or other mechanisms leading to multiparticle correlations.
\item (Mini)jets just produce clusters, so it is impossible to prove or disprove their existence based 
solely on the centrality dependence of the correlation data at soft/medium values of $p_T$.
\item Resonance decays in a thermal model yield a very small value of the $p_T$-covariance. 
\item In a model where matter forms lumps moving at similar collective velocity and 
particles move thermally within a cluster, description of the RHIC data requires 
on the average 6-15 particles in the cluster. In general, a larger number of particles within a
cluster helps to obtain the large (compared to $\sigma_p^2)/{\bar n}$) measured value of $\sigma_{\rm dyn}^2$.
\item In the ``lumped cluster'' model the $p_T^{\max}$ dependence of $F_{p_T}$
grows monotonically and then saturates. It would be interesting to confront this finding to the data.
\item Detailed microscopic modeling would be very useful in order to better 
understand the problem on transverse momentum correlations in relativistic heavy-ion collisions.

\end{enumerate}

\appendix

\section{Relation of measures of fluctuations to covariance}

One difficulty in comparing results of event-by-event 
momentum fluctuations 
presented by various experimental groups is the multitude of measures used.
Here we briefly show how under assumptions of 1)~small dynamical compared to
statistical fluctuations and 2)~sharp distribution in the multiplicity variable these
measures are simply proportional to the {\em covariance}. Although these remarks
are perhaps obvious to practitioners in the field, they seem worth
reminding, as discussions showed confusion.

Suppose we have events of class $n$ (formally, 
this can be any number chracteristic of the event: 
the multiplicity of detected particles, the number of participants, 
the response of a given detector, {\em etc.}, distributed according to 
the probability distribution $P_n$. 
Let $\rho_n(p_1,\dots,p_n)$ denote the
$n$-particle distribution of variables $p_i$ {\em within events of class $n$}, 
{\em e.g}, the distribution of transverse momenta in events of a fixed multiplicity $n$).
The subscript ${}_n$ indicates that $\rho$ depends functionally on $n$. The full 
probability distribution of obtaining event of class $n$ with momenta $p_1,\dots,p_n$ is
\begin{equation}
P_n\rho_n(p_1,\dots,p_n). 
\end{equation}
The {\em marginal} probability distributions are obtained from $\rho_n(p_1,\dots,p_{n})$
by integrating over $k$ momenta,
\begin{equation} 
\rho_n(p_1,\dots,p_{n-k}) = \int dp_{n-k+1}\dots dp_n \rho_n(p_1,\dots,p_n). 
\end{equation}
Next, we introduce the relevant moments for the distributions of class $n$:
\begin{eqnarray}
\overline{p_n}&=& \int dp \rho_n(p) p, \nonumber \\
\sigma^2_n(p)&=& \int dp \rho_n(p) (p-\overline{p_n})^2 , \nonumber \\
{\rm cov}_n(p_1,p_2)&=& \int dp_1 dp_2 \rho_n(p_1,p_2)(p_1-\overline{p_n})(p_2-\overline{p_n}),
\label{eq:cov} 
\end{eqnarray}
where $\rho_n(p)$ and $\rho_n(p_1,p_2)$ are the one- and two-particle {\em marginal} distributions 
within the class $n$.   
Now, in a typical setup we are interested in broader classes, containing $n$ in the range
$n_1 \le n \le n_2$. We denote for brevity $\sum_n=\sum_{n=n_1}^{n_2}$. 

Let us illustrate the basic statistical facts on the example of the measures 
$\sigma^2_{\rm dyn}$ and $F_{p_T}$. 
For other measures the analysis is analogous. Consider 
the variable $M_n=(p_1+\dots+p_n)/n$, {\em i.e.} the average value of the variable $p$. 
Then
\begin{eqnarray}
\overline{M}&=&\sum_n P_n \int dp_1 \dots dp_n \rho_n(p_1,\dots,p_n)\frac{p_1+\dots+p_n}{n}
=\sum_n P_n\overline{p_n}, \nonumber \\
\overline{M^2}&=&\sum_n P_n \int dp_1 \dots dp_n \rho_n(p_1,\dots,p_n)\frac{1}{n^2}
\sum_{i,j=1}^n \left [ (p_i-\overline{p_n})(p_j-\overline{p_n}) + \overline{p_n}^2 \right ], \nonumber \\
{\sigma^2_M}&=&\overline{M^2} - \overline{M}^2= \sum_n P_n \overline{p_n}^2 
- \left ( \sum_n P_n \overline{p_n} \right )^2 + \sum_n P_n \frac{\sigma^2_n(p)}{n}
+ \sum_n P_n \frac{1}{n^2} \sum_{i \neq j=1}^n {\rm cov}_n(p_i,p_j). \label{eq:exact}  
\end{eqnarray}
Suppose mixing of events is performed. Then, by definition, no correlations 
are present, {\em i.e.}, \mbox{${\rm cov}^{\rm mix}_n(p_i,p_j)=0$},  and 
\begin{equation} 
{\sigma^{2,{\rm mix}}_M} = \sum_n P_n \overline{p_n}^2 
- \left ( \sum_n P_n \overline{p_n} \right )^2 + \sum_n P_n \frac{\sigma^2_n(p)}{n}. \label{eq:mix}
\end{equation}
By definition, $\sigma^2_{\rm dyn}= \sigma^2_M -{\sigma^{2,{\rm mix}}_M}$. With above results
\begin{eqnarray}
\sigma^2_{\rm dyn} = \sum_n P_n \frac{1}{n^2}\sum_{i \neq j}{\rm cov}_n(p_i,p_j).
\end{eqnarray}
Note that all above results are {\em exact}, just following from obvious manipulations.
Thus $\sigma^2_{\rm dyn}$ is a {\em weighted sum of total (summed over particle pairs) 
covariances at fixed} $n$, $\sum_{i \neq j}{\rm cov}_n(p_i,p_j)$, with weights equal to 
$P_n \frac{1}{n^2}$. Moreover, all quantities in Eq.(\ref{eq:exact}) or (\ref{eq:mix}) are
possible to obtain experimentally from the given event sample. 

Now let us have a look on $F_{p_T} \equiv ( \sqrt{\omega} - \sqrt{\omega_{\rm mix}} )/\sqrt{\omega_{\rm mix}}$,
where $\omega = \sigma^2_M /\overline{M}$. We find immediately
\begin{equation}
F_{p_T}=\frac{\sigma_M}{\sigma^{{\rm mix}}_M } -1=\sqrt{1+\frac{\sigma^2_{\rm dyn}}
{\sigma^{2,{\rm mix}}_M }}-1. 
\end{equation}    
Again, this is an exact relation. At RHIC 
$\sigma^2_{\rm dyn} \ll \sigma^{2,{\rm mix}}_M$, hence we can
expand  
\begin{equation}
F_{p_T} \simeq \frac{1}{2} \frac{\sigma^2_{\rm dyn} }{\sigma^{2,{\rm mix}}_M},
\end{equation}
which shows the proportionality of the two measures in the limit of small dynamical correlations.

A further simplification occurs when the distributions $P_n$ are sharply peaked 
around some $\overline{n}$, which again is sufficiently well satisfied at RHIC. 
Then for a smooth function $f(n)$ 
\begin{equation} 
\sum_n \frac{P_n}{n^z} f(n) \simeq \frac{f(\overline{n})}{\overline{n}^z}. 
\end{equation}
In this sharp limit 
\begin{eqnarray}
\sigma^2_{\rm dyn} &\simeq& \frac{1}{\overline{n}^2} \sum_{i \neq j}{\rm cov}_{\overline{n}}(p_i,p_j),
\nonumber \\
\sigma^{2,{\rm mix}}_M &\simeq& \frac{\sigma^2_{\overline{n}}(p)}{\overline{n}}.
\end{eqnarray}

For other measures the results are similar. For $\Sigma^2_{p_T}$ we have from definition 
$\Sigma^2_{p_T} \equiv {\sigma^2_{\rm dyn}}/{\overline{p}^2}$, for $\Phi_{p_T}$ under conditions 1) and 2)
\begin{eqnarray}
\Phi_{p_T} \equiv \sqrt{\frac{\sigma_S^2}{\overline{n}}}-
\sigma_{\overline{n}}(p) \simeq \frac{\sum_{i \neq j} {\rm cov}_{\overline{n}}} 
{2 \overline{n}\sigma_{\overline{n}}(p)},
\end{eqnarray}
where $S_n= p_1+\dots+p_n$.  

{\bf Conclusion: Since at RHIC conditions 1) and 2) hold, the popular measures of event-by-event 
fluctuations are proportional to the covariance. Full information on correlations could be acquired 
by simply evaluating the covariance $\sum_{i \neq j} {\rm cov}_n$ for each $n$.} If 1) or 2) are relaxed, 
then, of course, the measures are no longer equivalent, but they are still related to the 
sum of the weighted covariances at various $n$ in the way dependent on the particular measure.  

Finally, we make a digression concerning the {\em inclusive} distributions, not used in our derivations but 
appearing frequently in similar studies. They should not be confused with the marginal distributions, 
to which they are related as follows:
\begin{eqnarray}
\rho_{\rm in}(p) &=& \sum_n P_n \int dp_1 \dots dp_n \sum_{i=1}^n \delta(p-p_i)
\rho_n(p_1,\dots,p_n), \nonumber \\
\rho_{\rm in}(p,q) &=& \sum_n P_n \int dp_1 \dots dp_n \sum_{i \neq j =1}^n 
\delta(p-p_i)\delta(q-p_j)\rho_n(p_1,\dots,p_n), 
\end{eqnarray}
with the properties 
\begin{eqnarray}
\int dp \rho_{\rm in}(p) &=& \overline{n}, \nonumber \\
\int dp dq \rho_{\rm in}(p,q) &=& \overline{n(n-1)}, \nonumber \\ 
\overline{p_{\rm in}}&=&\sum_n n P_n \overline{p_n}.
\end{eqnarray}

\end{document}